%% file: main.tex
\documentclass[sigconf]{acmart}

\AtBeginDocument{%
  }

\setcopyright{acmlicensed}
\copyrightyear{2018}
\acmYear{2018}
\acmDOI{XXXXXXX.XXXXXXX}
\acmConference[Conference acronym 'XX]{Make sure to enter the correct
  conference title from your rights confirmation email}{June 03--05,
  2018}{Woodstock, NY}
\acmISBN{978-1-4503-XXXX-X/2018/06}

\usepackage{amsmath}
\usepackage{graphicx}
\usepackage{booktabs}
\usepackage{multirow}
\usepackage{enumitem}

\begin{document}

\title{STORE: Semantic Tokenization, Orthogonal Rotation and Efficient Attention for Scaling Up Ranking Models}


\author{Yi Xu}
\affiliation{
  \institution{Alibaba Group}
  \city{Beijing}\country{China}
}
\email{xy397404@alibaba-inc.com}

\author{Chaofan Fan}
\affiliation{
  \institution{Alibaba Group}
  \city{Beijing}\country{China}
}
\email{fanchaofan.fcf@alibaba-inc.com}

\author{Jinxin Hu}
\affiliation{
  \institution{Alibaba Group}
  \city{Beijing}\country{China}
}
\email{jinxin.hjx@alibaba-inc.com}
\authornote{Corresponding author}

\author{Yu Zhang}
\affiliation{
  \institution{Alibaba Group}
  \city{Beijing}\country{China}
}
\email{daoji@alibaba-inc.com}

\author{Xiaoyi Zeng}
\affiliation{
  \institution{Alibaba Group}
  \city{Beijing}\country{China}
}
\email{yuanhan@taobao.com}

\author{Jing Zhang}
\affiliation{
  \institution{Wuhan University}
  \city{Wuhan}\country{China}
}
\email{jingzhang.cv@gmail.com}
\renewcommand{\shortauthors}{Trovato et al.}
\vspace{-0.3cm}

\begin{abstract}
Ranking models have become an important part of modern personalized recommendation systems. However, significant challenges persist in handling high-cardinality, heterogeneous, and sparse feature spaces, particularly regarding model scalability and efficiency.
We identify two key bottlenecks: 
(i) Representation Bottleneck: Driven by the high cardinality and dynamic nature of features, model capacity is forced into sparse-activated embedding layers, leading to low-rank representations. This, in turn, triggers phenomena like "One-Epoch" and "Interaction-Collapse," ultimately hindering model scalability.
(ii) Computational Bottleneck: Integrating all heterogeneous features into a unified model triggers an explosion in the number of feature tokens, rendering traditional attention mechanisms computationally demanding and susceptible to attention dispersion.
To dismantle these barriers, we introduce STORE, a unified and scalable token-based ranking framework built upon three core innovations:
(1) Semantic Tokenization fundamentally tackles feature heterogeneity and sparsity by decomposing high-cardinality sparse features into a compact set of stable semantic tokens; and
(2) Orthogonal Rotation Transformation is employed to rotate the subspace spanned by low-cardinality static features, which facilitates more efficient and effective feature interactions; and 
(3) Efficient attention that filters low-contributing tokens to improve computional efficiency while preserving model accuracy. Across extensive offline experiments and online A/B tests,  our framework consistently improves prediction accuracy(online CTR by 2.71\%, AUC by 1.195\%) and training effeciency (1.84× throughput).
\end{abstract}
\vspace{-0.8cm}
\vspace{-0.3cm}
\vspace{-0.1cm}
\maketitle
\vspace{-0.5cm}
\section{Introduction}
\input{sects/intro_v2}

\vspace{-0.3cm}
\section{Methodology} 
\input{sects/method_v2}
\vspace{-0.1cm}
\section{Experiments}
\input{sects/exp_format_v2}
 \vspace{-0.15cm}
\section{Online Experiments}
We conducted a 15-day online A/B test on a large-scale e-commerce platform, comparing STORE with the production baseline. STORE achieved a relative CTR increase of 2.71\%. In deployment, the OPMQ is set to K=32, codebook size=300 for SIDs. The sparsity of attention is set to 1/2, improving inference efficiency and response speed while maintaining performance.
\section{Conclusion}
We introduced STORE, a framework that resolves both representation and computational bottlenecks in recommenders. Through semantic tokenization, orthogonal rotation, and an efficient attention mechanism, STORE unlocks superior model scalability and computational efficiency. Extensive experiments validate STORE as a practical and effective path toward building more powerful large-scale ranking models.
\vspace{-0.1cm}
\bibliographystyle{ACM-Reference-Format}
\bibliography{main}
\end{document}

%% file: sects/intro_v2.tex
Ranking models form the backbone of modern online services. Their core task is to model complex user behavior by processing a vast and heterogeneous collection of features. To handle this feature diversity, existing ranking models have evolved into a collection of specialized modules for feature interaction. While effective for accuracy, this intricate and fragmented design presents a major obstacle to scalability. Unlike Large Language Models (LLMs), where "Scaling Laws" provide a clear path to predictable performance gains, ranking models fail to exhibit similar scaling behavior\cite{scale:journals/corr/abs-2001-08361,HSTU}. There are two fundamental challenges that prevent ranking models from benefiting from Scaling Laws:

(1) Representation Bottleneck: The high-cardinality features force model capacity into sparse-activated embedding layers over deep networks. This yields low-rank embeddings, triggers the "One-Epoch"\cite{one_epoch} and "Interaction-Collapse"\cite{collapse} problems. Ultimately, this hinders model scalability where adding depth or training epochs offers diminishing returns.
This phenomenon leads to the loss of effective high-order feature interactions\cite{collapse}. As model size grows, gains diminish rapidly, undermining capacity utilization and predictable scaling. 
(2) Computational Bottleneck: As the model scale to incorporate vast feature sets triggers an explosion in the number of feature tokens. This renders vanilla attention, with its $O(L^{2})$ complexity, computationally prohibitive and simultaneously exacerbates attention dispersion, where vital signals are lost amidst a sea of irrelevant tokens.

\begin{figure*}[htbp]
\vspace{-0.3cm}
    \centering
    \includegraphics[width=0.7\textwidth]{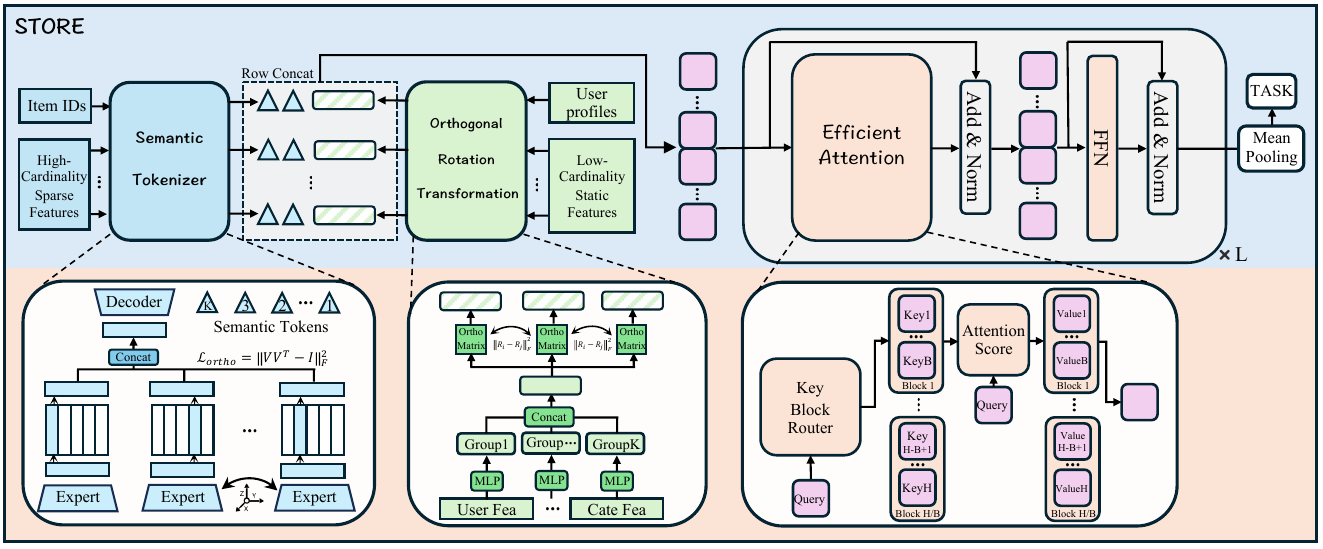}
    \caption{Overview of the proposed STORE.}
    \label{fig:method_overview}
    \vspace{-0.5cm}
\end{figure*}
To address these challenges, we introduce STORE (\textbf{S}emantic \textbf{T}okenization, \textbf{O}rthogonal \textbf{R}otation, and \textbf{E}fficient attention), a unified and scalable ranking framework built upon three synergistic components.
(1) Semantic Tokenization efficiently decomposes high-cardinality features into a set of compact, orthogonal semantic IDs. This approach fundamentally mitigates feature heterogeneity and sparsity, thereby unlocking more efficient model scaling.
(2) Orthogonal Rotation Transformation is employed to rotate the subspace spanned by low-cardinality static features, which facilitates more efficient and effective feature interactions in different high-dimensional spaces. 
(3) Efficient Attention adaptively utilizes sparsity mechanism to prune low-contributing tokens conditioned on the target item and context, reducing computational complexity and alleviating attention dispersion while preserving accuracy. 
Our key contributions are summarized as follows:
\begin{itemize}[topsep=0pt, partopsep=0pt]
\item We present STORE, a unified token‑based ranking model framework that effectively tackles heterogeneity and sparsity in large and dynamic feature spaces. This paradigm mitigates long‑standing scaling‑law bottlenecks in large‑scale recommender systems, supporting more predictable scaling.
\item We design a synergistic trio of architectural innovations: semantic tokenization and orthogonal rotation are proposed to resolve the representation bottleneck, while efficient attention is proposed to mitigate the computational bottleneck by sparsifying attention and reducing quadratic costs.
\item Extensive offline experiments and online A/B tests demonstrate the superiority of STORE in both effectiveness and efficiency, showing an improvement of 1.195\% in AUC, 2.71\% in CTR, and 1.84× higher training throughput.
\end{itemize}

%% file: sects/method_v2.tex
This section details the proposed STORE framework. 
To address the heterogeneity and sparsity of feature space, features are categorized into high-cardinality sparse features(e.g., item identifiers) and low-cardinality static features(e.g., category id, age, gender), employed wtih distinct processing strategies: Semantic Tokenizer for high-cardinality sparse features, detailed in section \ref{sec:semantic_tokenizer}, and Orthogonal Rotation Transformation for low-cardinality static features, detailed in section \ref{sec:Orthogonal_Rotation}.
To migrate the computational efficiency bottleneck for ranking model, the Efficient Attention for Unified Feature Interaction is proposed, detailed in section \ref{sec:attention}.
\vspace{-0.3cm}
\subsection{Semantic Tokenizer}
\label{sec:semantic_tokenizer}
Using item IDs as a representative example of high-cardinality sparse features, we mitigate this issue by mapping raw item IDs into a more stable and structured semantic space via Semantic IDs (SIDs). We achieve this by quantizing powerful pre-trained embeddings (e.g., from SASRec) into a sequence of SIDs.
\begin{equation}
\vspace{-3pt}
\label{eq:item_tokenization}
  (SID_1, SID_2, \dots, SID_K) = \mathcal{T}_{\text{item}}(\mathbf{e}_{p} \in \mathbb{R}^{d})
\end{equation}
where $\mathbf{e}_{p}$ denotes the pre-trained item embedding and $\mathcal{T}_{\text{item}}$ denotes the item semantic tokenization function producing $K$ SIDs for each item. In this paper, the setting is $K=H$. 

To efficiently encode high-cardinality IDs into a set of compact and parallel SIDs, we propose an \textbf{O}rthogonal, \textbf{P}arallel, \textbf{M}ulti-expert \textbf{Q}uantization network(\textbf{OPMQ}). For each item, the network utilizes $K$ experts to encode its pre-trained embedding into $K$ latent representations. The formulation is as follows.
\begin{align}
    \mathbf{z}_{i} &= E_{i}(\mathbf{e}_{p}), \quad i \in \{1, \dots, K\} \\
    c_i &= \arg\min_{j \in \{1, \dots, V\}} \| \mathbf{z}_i - \mathbf{s}_{j} \|_2^2
\end{align}
where $E_{i}$ is the $i$-th expert and the latent representation is $\mathbf{z}_{i}$. For each latent vector $\mathbf{z}_i$, we assigning it to the index of its nearest neighbor codeword $c_i$, the codeword vector $\mathbf{s}_i$. The entire OPMQ network is trained end-to-end by minimizing the reconstruction error between the original embedding and the output of a decoder that aggregates the quantized vectors.
\vspace{-3pt}
\begin{equation}
    \mathcal{L}_{recon}=||\mathbf{e}_{p}-decoder[\sum_i^K(\mathbf{z_i}+sg(\mathbf{s}_{i}-\mathbf{z_i}))]||^2
\end{equation}
To capture diverse and non-redundant aspects of the original item, the orthogonal regularization of SIDs is performed on the parameters of multi-experts. Formally, for the $i$-expert, we define the parameter vector $\mathbf{V_i} \in \mathbb{R}^{d_1d_2}$ as the L2-normalized version of the flattened parameter matrix $\mathbf{W}_i \in \mathbb{R}^{d_1 \times d_2}$.
The orthogonal regularization is performed on the set of $K$ parameter vectors, which is formulated as follows.
\vspace{-3pt}
\begin{equation}
\mathcal{L}_{\text{orth}} = \left\| \mathbf{V} \mathbf{V}^\top - \mathbf{I} \right\|_F^2,
\end{equation}
where \( \mathbf{I} \) is the identity matrix and \( \|\cdot\|_F^2 \) denotes the Frobenius norm.
\vspace{-0.5cm}
\subsection{Orthogonal Rotation Transformation}
\label{sec:Orthogonal_Rotation}
Unlike high-cardinality sparse features with  heterogeneity and sparsity, for low-cardinality static features with controllable features sizes, we employ the original embeddings. For simplicity and efficiency, we perform manual grouping based on their semantic meanings of domain knowledge. These features are partitioned into $K$ semantic groups, with each group containing several features. For each feature group, a shallow MLP is employed for intra-group feature fusion. By concatenating all the semantically fused feature groups, we obtain an instance-wise feature block, denoted as $\mathbf{C}$. The formulation is as follows.
\vspace{-3pt}
\begin{equation}
    \mathbf{C}=[MLP_1(g_1),\dots,MLP_K(g_K)]
\end{equation}
To facilitate efficient and effective feature interactions in high-dimensional spaces, the orthogonal rotation transformation is employed to rotate the instance-wise feature block. To obtain $K$ diverse instance-wise feature block, we rotate the $\mathbf{C}$ with $K$ group of orthogonal matrices. For the $i$-th rotation, the formulation is as follows.
\begin{equation}
\vspace{-3pt}
    \mathbf{O_i} = \mathbf{C}\mathbf{R_i}
\end{equation}
where $\mathbf{R_i}$ is an orthogonal matrix. To prevent the rotation matrices from collapsing during training (e.g., all becoming identical), we introduce a diversity regularization term. This works in concert with the orthogonality constraint to encourage a diverse set of learned transformations, formulated as follows.
\vspace{-3pt}
\begin{align}
    \min_{\mathbf{R}_1, \dots, \mathbf{R}_k} \quad & -\lambda \sum_{i=1}^{K} \sum_{j=i+1}^{K} \| \mathbf{R_i} - \mathbf{R_j} \|_F^2 \\
    \text{s.t.} \quad & \mathbf{R_i}^T\mathbf{R_i} = \mathbf{I}, \quad \forall i \in \{1, \dots, K\}
\end{align}
\vspace{-3pt}
where $||\cdot||_F$ is the Frobenius norm. $\lambda$ is the hyperparameter, which is set to 0.1 in this paper. The rotation matrices and the parameters of the main network are alternative optimized.
\vspace{-0.1cm}
\subsection{Efficient Attention for Unified Feature Interaction}
\label{sec:attention}
To efficiently capture feature interactions in a unified framework, we propose the efficient attention with instance-wise tokens. Following the distinct processing of high-cardinality sparse and low-cardinality static features, we concatenate the embedding of SIDs with the rotated feature block in the first layer, $\mathbf{\mathbf{X_0^i}}=[\mathbf{s_i},\mathbf{O_i}],\mathbf{X_0^i} \in \mathbb{R}^{H \times d}$. 
The input of efficient attention for feature interactions is the instance-wise token sequence $\mathbf{X_0=[X_0^1,X_0^2,\dots,X_0^H]}$,  which construct the $\mathbf{Q},\mathbf{K},\mathbf{V}$. 

The iterative unified efficient attention for feature interaction formulated as follows:
\vspace{-1pt}
\begin{align}
\mathbf{X_{l}} &= \text{LN}(\text{EfficientAttention}(\mathbf{X_{l-1}}) + \mathbf{X_{l-1}}) \label{eq:sparse_attention} 
\end{align}
where $\mathbf{X_{l-1}}$ is the input to the $\mathbf{l}$-th layer, $\text{LN}$ denotes Layer Normalization. Vanilla self-attention's $O(H^{2})$ computational complexity becomes prohibitive as the number of instance-wise tokens $H$ grows.
\vspace{-0.1cm}
\begin{align}
\vspace{-0.4cm}
    \text{MoBA}(\mathbf{Q}, \mathbf{K}, \mathbf{V}) &= \text{Softmax}\left(\mathbf{Q}\mathbf{K}[Ind]^T\right)\mathbf{V}[Ind], \label{eq:moba_attention_formula_bold} \\
    Ind_i &= [(i-1) \times B + 1, i \times B] \label{eq:block_indices} 
\end{align}
To overcome this bottleneck, our framework incorporates an efficient attention mechanism. Specifically, we adopt MOBA~\cite{moba}, which employs a routing strategy for each query to attend to only a small subset of key-value pairs. As formulated in Eq \ref{eq:moba_attention_formula_bold},\ref{eq:block_indices}, $Ind_i \subseteq \{1, \dots, H\}$  is the dynamically selected set of indices of key-value pairs, the size of selective block is $B$. This approach reduces the complexity from quadratic significantly, a choice made viable by our framework's effective mitigation of feature heterogeneity and sparsity.
\vspace{-6pt}

%% file: sects/exp_format_v2.tex
In this paper, we conduct extensive experiments on both industrial and public datasets to evaluate the effectiveness of STORE with the following questions: 

\textbf{RQ1}: How does STORE compare to SOTA ranking models?

\textbf{RQ2}: What is the contribution of each component in STORE? 

\textbf{RQ3}: What is the scalability of STORE?
\vspace{-0.2cm}
\subsection{Experimental Setup}
\begin{table}[htbp] %
\vspace{-0.3cm}
  \centering
  \caption{Overall performance comparison in CTR prediction on public and industrial datasets. "Improv." denotes the relative improvement of STORE over the best baseline. The best baseline performance score is denoted in \underline{underline}.}
  \label{tab:performance_overall}
  \resizebox{\columnwidth}{!}{%
    \begin{tabular}{llcccccc} 
      \toprule
      \multicolumn{2}{c}{\textbf{Dataset}} & \multicolumn{3}{c}{\textbf{Avazu}} & \multicolumn{3}{c}{\textbf{Industrial}} \\
      \cmidrule(lr){3-5} \cmidrule(lr){6-8}
      \multicolumn{2}{c}{\textbf{Method}} & \textbf{AUC} & \textbf{GAUC} & \textbf{Logloss} & \textbf{AUC} & \textbf{GAUC} & \textbf{Logloss} \\
      \midrule
       \multicolumn{2}{l}{FM} & 0.7291 & 0.7248 & 0.4052 & 0.6711 & 0.6011 & 0.1144 \\
       \multicolumn{2}{l}{DNN}  &0.7231 &0.7211 &0.4052 & 0.6721 & 0.6005 & 0.1148 \\
       \multicolumn{2}{l}{Wide\&Deep} & 0.7356 & 0.7329 & 0.3988 & 0.6720 & 0.6018 & 0.1144 \\
       \multicolumn{2}{l}{DeepFM} & 0.7404 & 0.7375 & 0.3965 & 0.6707 & 0.5907 & 0.1152 \\
       \multicolumn{2}{l}{DCN} & 0.7344 & 0.7310 & 0.4042 & 0.6734 & 0.6029 & 0.1141 \\
       \multicolumn{2}{l}{AutoInt} & 0.7439 & 0.7408 & 0.3948 & 0.6728 & 0.6021 & 0.1142 \\
       \multicolumn{2}{l}{GDCN} & 0.7370 & 0.7344 & 0.3989 & 0.6726 & 0.6022 & 0.1142 \\
      \multicolumn{2}{l}{MaskNet} & 0.7426 & 0.7383 & 0.3942 & 0.6753 & 0.6054 & \underline{0.1140} \\
      \multicolumn{2}{l}{PEPNet} & 0.7411 &0.7380 &0.3961 & 0.6741 & 0.6039 & 0.1148 \\
      \multicolumn{2}{l}{RankMixer} & 0.7450 & 0.7412 & 0.3951 & \underline{0.6774} & 0.6053 & \underline{0.1140} \\
     \multicolumn{2}{l}{OneTrans} & \underline{0.7461} & \underline{0.7432} & \underline{0.3943} & 0.6771 & \underline{0.6058} & 0.1141 \\
      \midrule
      \multicolumn{2}{l}{\textbf{STORE}} & \textbf{0.7479} & \textbf{0.7451} & \textbf{0.3912} & \textbf{0.6804} & \textbf{0.6064} & \textbf{0.1139} \\
      \multicolumn{2}{l}{\textbf{STORE-4 Epoch}} & \textbf{0.7488 }& \textbf{0.7463} & \textbf{0.3900} & \textbf{0.6855} & \textbf{0.6086} & \textbf{0.1134} \\
      \multicolumn{2}{l}{\textbf{Improv.}} &\textit{ +0.362\%} & \textit{+0.417\%} &\textit{ +0.913\% }& \textit{+1.195\%} &\textit{+0.462\%} & \textit{+0.526\%} \\
      \bottomrule
    \end{tabular}%
  }
  \vspace{-0.6cm}
\end{table}
\subsubsection{Dataset}

To validate the effectiveness of our proposed framework, we conduct experiments on real-world large-scale datasets and public datasets. 

\textbf{Avazu}: Avazu is a widely-used public benchmark for CTR
prediction, consisting of 9 million of chronologically ordered ad click logs, 23 feature fields and 3437 site ids.

\textbf{Industrial Dataset}: This dataset contains 7 billion user interaction records from an international e-commerce advertising system, featuring diverse item features and  user behavior sequences.
\vspace{-6pt}
\subsubsection{Evaluation Metrics} For the evaluation, we use the widely used AUC, GAUC and LogLoss for prediction accuracy. 
We use training TFlops/Batch(batch size=1024) to evaluate the training effeciency.
\vspace{-6pt}
\sloppy
\subsubsection{Baselines} To demonstrate the effectiveness of our framework, we evaluate the proposed framework with state-of-the-art CTR prediction models, including FM\cite{FM}, DNN, Wide\&Deep\cite{widedeep}, DeepFM\cite{DeepFM}, DCN\cite{dcn}, AutoInt\cite{AutoInt}, GDCN\cite{gdcn}, MaskNet\cite{masknet}, PEPNet\cite{PEPNet}, RankMixer\cite{rankmixer} and OneTrans\cite{onetrans}.
\vspace{-6pt}
\subsubsection{Implementation Details} We utilize the pre-trained item embeddings from a pre-trained SASRec model. The number of SIDs (K) and the codebook size are set to (3, 16) for the public dataset and (32, 300) for the industrial dataset, respectively. 
\vspace{-0.2cm}
\subsection{Overall Performance(RQ1)} Table \ref{tab:performance_overall} presents the overall prediction performance of all methods on both industrial and public datasets, alongeside with the relative improvement against the best baseline. The performance comparision on two datasets can demonstrate the effectiveness of STORE. It is noteworthy that while methods like Rankmixer and OneTrans project aggregated feature groups to mitigate feature heterogeneity, achieving gains over traditional networks.
In contrast, our proposed method fundamentally resolves the issues by leveraging SIDs and rotation techniques. This leads to substantial improvements in prediction accuracy.
\vspace{-0.2cm}
\subsection{Ablation Study(RQ2)}
As shown in Table \ref{tab:tiger_ablation}, we conduct ablation experiments to evaluate the impact of each component in STORE. 
\begin{itemize}[topsep=0pt, partopsep=0pt]
\item \textbf{Different Semantic Tokenizer}: 
We compare OPMQ against two widely-used tokenizers: RQ-VAE\cite{rqvae}, and Optimized Product Quantization (OPQ)\cite{opq}. 
\item \textbf{Orthogonal Rotation Transformation}: We conduct experiments on w/o orthogonal rotation transformation.
\item \textbf{Efficient Attention}: Compared with vanilla attention, the efficient attention achieve comparable prediction accuracy with higher training effeciency.
\end{itemize}
 \vspace{-0.1cm}
\begin{table}[htbp]
  \centering
  \caption{Ablation study of STORE variants and components.}
  \vspace{-0.3cm}
  \label{tab:tiger_ablation}
  \resizebox{\columnwidth}{!}{%
  \begin{tabular}{lcccc}
    \toprule
    \multicolumn{1}{l}{\textbf{Variants}} & \textbf{AUC} & \textbf{GAUC} & \textbf{Logloss}& \textbf{TFlops/Batch} \\
    \midrule
      \textbf{STORE-4 Epoch} & \textbf{0.6855} & \textbf{0.6086} & \textbf{0.1134}& \textbf{1.764} \\
      \textbf{STORE} & \textbf{0.6804} &\textbf{ 0.6064} & \textbf{0.1139}& \textbf{1.763} \\
       w OPQ & 0.6787 & 0.6045 & 0.1140 & 1.763 \\
       w RQ-VAE & 0.6768 & 0.6047 & 0.1141& 1.762 \\
       w/o Orthogonal Rotation & 0.6780 & 0.6050 & 0.1140& 1.760 \\
       w Vanilla-Attention & 0.6812 & 0.6068 & 0.1137 & 3.240 \\
   
    \bottomrule
  \end{tabular}
  }
\vspace{-0.3cm}
\end{table}

\vspace{-0.2cm}
\subsection{Scaling Laws Study with Different Hyperparameters(RQ3)}
\begin{figure}[htbp] 
\vspace{-0.3cm}
    \centering
    \includegraphics[width=0.9\linewidth]{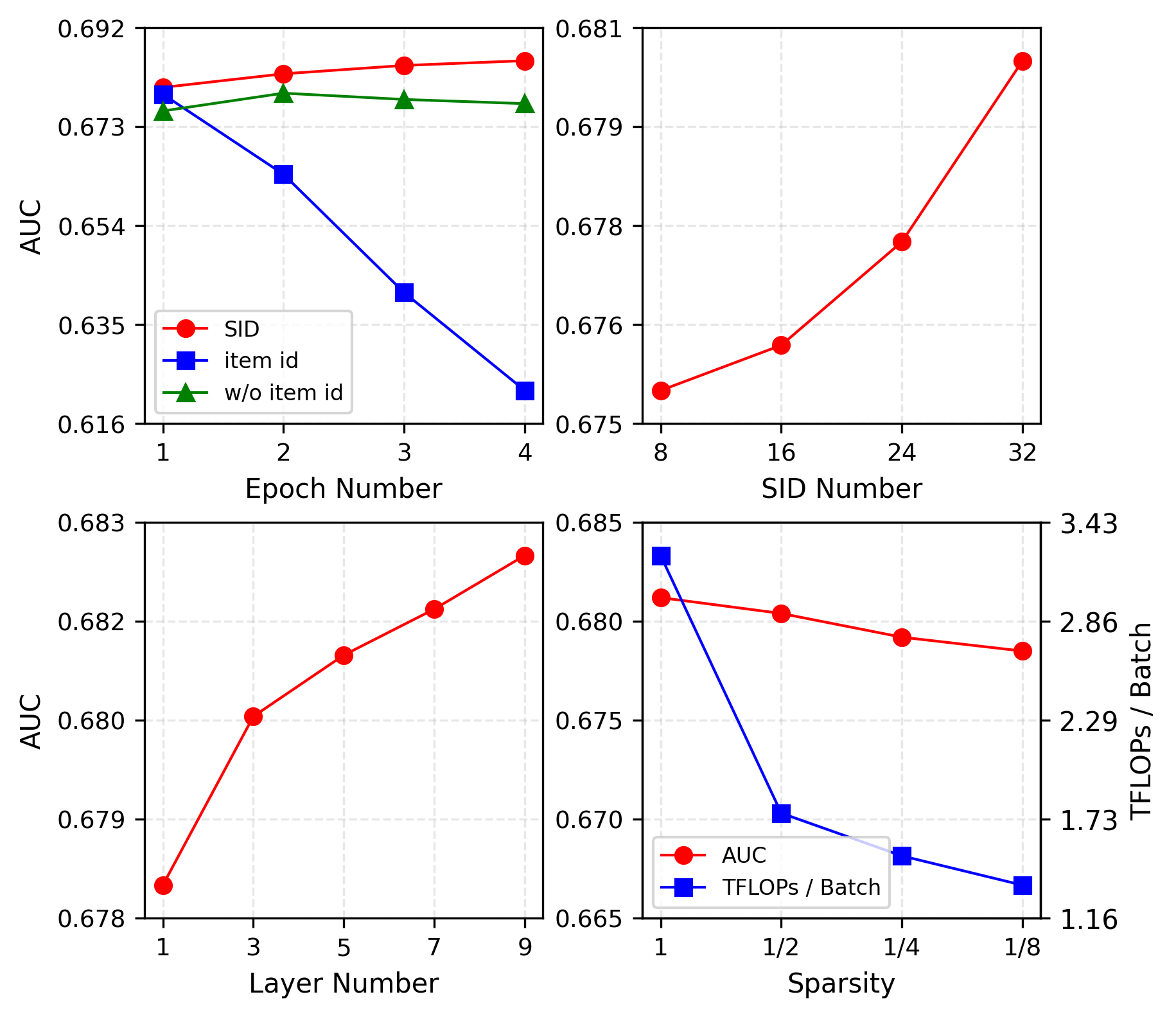}
    \caption{Scaling Laws Study of (a) Epoch Number (b) SID Number (c) Layer Number (d) Sparsity.}
    \label{fig:hyper_exp}
\vspace{-0.5cm}
\end{figure}

In this section, we conducted experiments to demonstrate the scalability and efficiency of our proposed method. We will compare them from the following dimensions. As shown in Fig \ref{fig:hyper_exp}: (a) Epoch number: that demonstrate SIDs combats the "One-Epoch" phenomenon which models with ItemIDs show decreased performance over mult-epoch. (b) SID Number: The more SIDs, the better the effect. (c) Layer Number: The more Layers, the better the effect. (d) Sparsity: We explore the relationship between attention sparsity, the training effeciency and model accuracy has demonstrated STORE's ability to reduce computational cost with minimal impact on performance.
